# A GPU-enabled finite volume solver for large shallow water simulations

Fabrice Zaoui[1], unpublished work (date: 2016 January)

*EDF R&D, LNHE, 6 quai Wattier, Chatou, 78400, France*

**Abstract**

This paper presents the implementation of a HLLC finite volume solver using GPU technology for the solution of shallow water problems in two dimensions. It compares both CPU and GPU approaches for implementing all the solver's steps. The technology of graphics and central processors is highlighted with a particular emphasis on the CUDA architecture of NVIDIA. The simple and well-documented Application Programming Interface (CUDA API) facilitates the use of the display card workstation as an additional computer unit to the central processor. Four professional solutions of the NVIDIA Quadro line are tested. Comparison tests between CPU and GPU are carried out on unstructured grids of small sizes (up to 10,000 elements), medium and large sizes (up to 10,000,000 elements). For all test cases, the accuracy of results is of the same order of magnitude for both approaches. Furthermore, the obtained speed gains with the GPU strongly depend on the model of the graphics card, the size of the problem and the simulation time.

## 1. Introduction

Models based on Shallow Water Equations (SWE) are useful in Computational Fluid Dynamics. They can simulate a wide range of geophysical flows like rivers, channels, estuarine and coastal circulations, dam breaks, etc. These phenomena occur in large domains and may require long simulation time for a real study. Consequently the computation times can be extremely high if an efficient solver is not implemented. Fortunately, the numerical solution of the 2D hyperbolic systems of conservation laws SWE can be easily implemented on parallel machines to keep the execution times acceptable. The parallelism is mainly concerned with the domain decomposition method on cluster, multithreaded programs and vectorization techniques.

Most of the codes solving 2D SWE in parallel are historically based on CPUs. Even if the GPU architecture is made for massive parallelism, its use to obtain the solution of non-graphics applications like mathematical problems in science and engineering is more recent, and requires a very different programming model from that of CPU [1-3]. When a code is developed and optimized to benefit from the large memory bandwidth and Floating Point Operations Per Seconds (FLOPS) of the GPU architecture, simulations can run significantly faster than a CPU model [4].

In this paper, NVIDIA Quadro cards are used to implement the solution of 2D SWE with the finite volume method on unstructured grids [5]. Performances are compared with a mono-core CPU version of the same algorithm.

## 2. Method

*2.1. Mathematical model*

The conservative form of 2D SWE is:

$$\frac{\partial \mathbf{U}}{\partial t} + \frac{\partial \mathbf{G(U)}}{\partial x} + \frac{\partial \mathbf{H(U)}}{\partial y} = \mathbf{S(U)} \quad in \ \Omega_t \times [0, T_s] \tag{1}$$

$$\mathbf{U} = \begin{bmatrix} h \\ hu \\ hv \end{bmatrix}, \mathbf{G} = \begin{bmatrix} hu \\ hu^2 + \frac{gh^2}{2} \\ huv \end{bmatrix}, \mathbf{H} = \begin{bmatrix} hv \\ huv \\ hv^2 + \frac{gh^2}{2} \end{bmatrix},$$

---

[1] Corresponding author : fabrice.zaoui@edf.fr

$$\mathbf{S} = gh(\mathbf{S_0} + \mathbf{S_f}), \mathbf{S_0} = \begin{bmatrix} 0 \\ S_{0x} = -\dfrac{\partial z}{\partial x} \\ S_{0y} = -\dfrac{\partial z}{\partial y} \end{bmatrix}, \mathbf{S_f} = \begin{bmatrix} 0 \\ S_{fx} = -\dfrac{n^2 u\sqrt{u^2+v^2}}{h^{4/3}} \\ S_{fy} = -\dfrac{n^2 v\sqrt{u^2+v^2}}{h^{4/3}} \end{bmatrix}$$

where $h = \eta - z$ is the water depth $(m)$, $\eta$ is the free surface $(m)$, $z$ is the bathymetry $(m)$, $u$ and $v$ are the velocity components in the $x$ and $y$ directions $(m/s)$, $g$ is the acceleration due to gravity $(m/s^2)$, $n$ is the Manning roughness coefficient $(s/m^{1/3})$ and $T_s$ is the simulation time $(s)$.

*2.2. Discretization*

The computational domain is divided into a set of triangles and the partial differential equation (1) is solved with a cell-centered finite volume method. This implies the discretization of the SWE in integral form:

$$\frac{\partial}{\partial t}\int_\Omega \mathbf{U}\, d\Omega + \int_\Omega \nabla \cdot \mathbf{F}\, d\Omega = \int_\Omega \mathbf{S}\, d\Omega \qquad (2)$$

where $\mathbf{F(U)} = (\mathbf{G(U)}, \mathbf{H(U)})^T$ and $\Omega$ denotes the volume of the cell $i$. The application of the divergence theorem for a triangular cell gives:

$$\int_\Omega \nabla \cdot \mathbf{F}\, d\Omega = \oint_\Gamma \mathbf{F(U)} \cdot \mathbf{n}\, d\Gamma = \sum_{k=1}^{3} \mathbf{F}_k(\mathbf{U}) \cdot \mathbf{n}_k l_k \qquad (3)$$

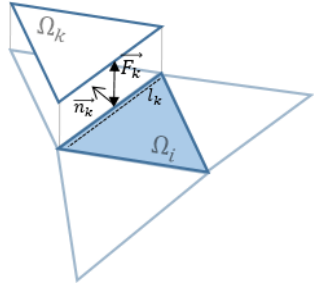

Fig. 1. Control volume $\Omega_i$ of cell-centered scheme.

Assuming the state vector $\mathbf{U}$ to be piecewise constant over cells and a simple explicit Euler scheme for the time discretization, the solution for a new time step $n+1$ is computed as:

$$\frac{\mathbf{U}_i^{n+1} - \mathbf{U}_i^n}{\Delta t} = -\frac{1}{\Omega_i}\sum_{k=1}^{3} \mathbf{F}_k(\mathbf{U}) \cdot \mathbf{n}_k l_k + \mathbf{S} \qquad (4)$$

The time step $\Delta t$ must be selected carefully to maintain the stability of the scheme. Therefore the Courant-Friedrichs-Lewy (CFL) stability condition is applied with a Courant number equal to $0.7$.

The HLLC approximate Riemann solver [6] is used to compute the numerical fluxes $\mathbf{F}_k(\mathbf{U})$ of mass and momentum at interfaces. Source terms are treated with a hydrostatic reconstruction for the topography $\mathbf{S_0}$ and a semi-implicit method for the friction $\mathbf{S_f}$.

*2.3. Implementation*

The implementation of algorithm (4) is written in the C programming language for CPU and duplicated in CUDA C for GPU. CPU has to perform in a wide range of applications. It is traditionally optimized for single-thread performance with a big cache memory, sophisticated control logic and high clock frequency. GPU is specialized for highly parallel tasks with its high number of processing cores optimized for a multithreaded execution.

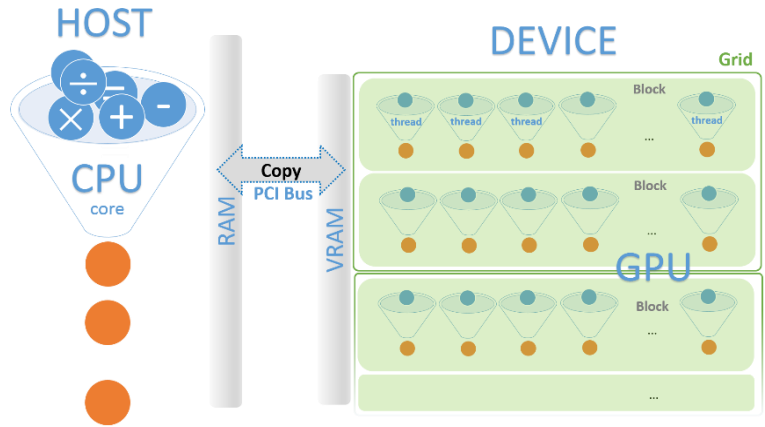

Fig. 2. CUDA source code can address HOST (CPU) and DEVICE (GPU).

The best performance for CUDA code is expected if the number of data transfer is minimized between HOST and DEVICE through the PCI express bus, see Figure 2. Consequently, the loop time is on the HOST with calls for calculation on DEVICE. Transfer copy is limited to the initial data structure from the input data (copy from HOST to DEVICE) and to output calculation results at $t = T_s$ (copy from DEVICE to HOST).

Functions to compute (4) are executed by many threads on GPU. The threads are grouped in grids of blocks that run in parallel to compute the numerical flux $\mathbf{F}_k(\mathbf{U})$ and to update the new state $\mathbf{U}^{n+1}$. These two steps are mainly concerned by a parallel loop over the number of interfaces and cells respectively.

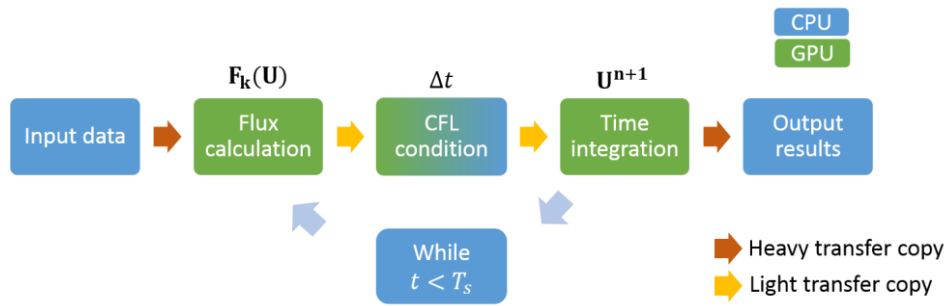

Fig. 3. Flowchart of parallel computing.

The algorithm of the finite volume solver is verified on a set of well-known benchmark cases. For example, Figure 4 shows the water depth evolution of a dam-break flooding over three mounds. This case considers a non-trivial wetting and drying process.

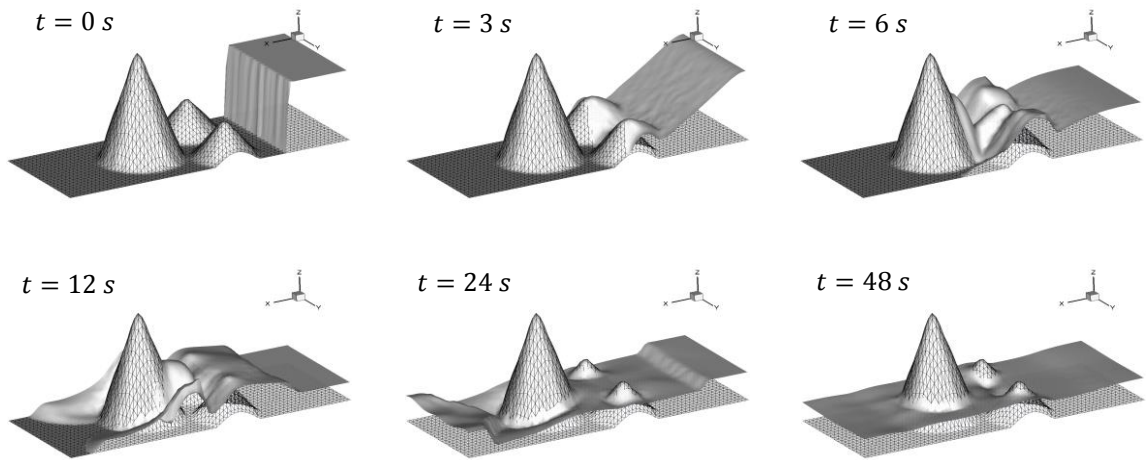

Fig. 4. Topography, mesh and water depth evolution of the three-mound problem.

## 3. Results of the comparison tests

Four cards of the NVIDIA Quadro line are tested, see Table 1. These GPUs are professional graphics solutions. They are compared with the INTEL Xeon CPUs working in a sequential mode. These CPUs are Xeon E3 3.2 GHz for the Quadro K600, Xeon E3 3.4 GHz for Quadro K2000, and model E5 2.6 GHz for K5200/K6000.

Table 1. GPU solutions for the tests.

| Model name | K600 | K2000 | K5200 | K6000 |
| --- | --- | --- | --- | --- |
| Number of processing cores | 192 | 384 | 2,304 | 2,880 |
| TeraFLOPS (Single precision, Peak) | 0.34 | 0.73 | 3.5 | 5.1 |
| Memory (GB) | 1 | 2 | 8 | 12 |
| Memory bandwidth (GBps) | 29 | 64 | 192 | 288 |

The test case is a water drop in a large basin. The fluid is initially at rest with a Gaussian free surface at the center of a large square domain with reflective conditions on all the boundaries. The longest studied simulation time $T_s$ is equal to 2,400 s.

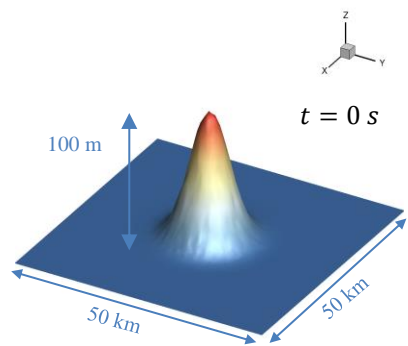

Fig. 5. Initial state of the water drop.

Comparison tests are carried out on unstructured grids of different sizes ranging from 1,000 to 10,000,000 cells. Table 2 gives information about sizes of the grids.

Table 2. Unstructured grids.

| Test number | 1 | 2 | 3 | 4 | 5 |
|---|---|---|---|---|---|
| Number of cells (triangles) | 1,036 | 10,132 | 104,788 | 1,056,518 | 10,261,932 |
| Number of edges | 1,588 | 15,310 | 157,542 | 1,585923 | 15,396,468 |
| Number of nodes | 553 | 5,179 | 52,755 | 529,406 | 5,134,537 |

The tests are run with a double precision representation. Final results for CPU and GPU at $t = T_s$ are verified as being very close.

The CPU/GPU ratios of computation times are shown in Figure 6 for the first four tests. As the necessary amount of memory required by the test number #5 is large, it can only be run on the GPUs K5200 and K6000 and is not shown in Figure 6.

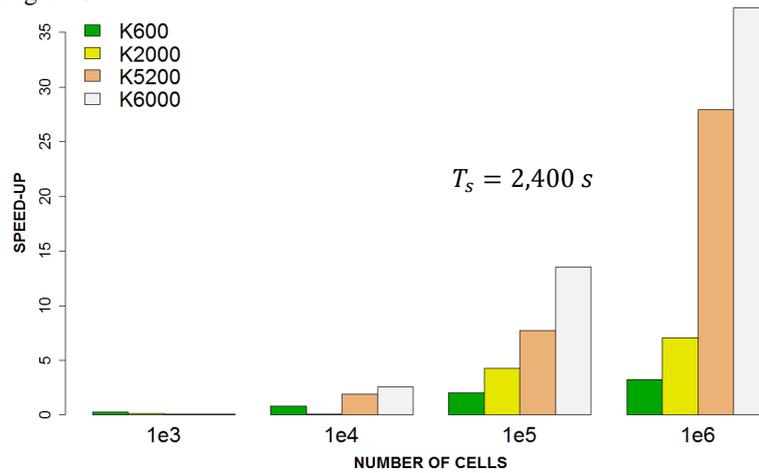

Fig. 6. Relative performance assessment on processors.

Considering the mean speed-up offered by each GPU for the first four tests, a ranking can be established as shown in Figure 7. This can be interpreted as a work distribution according to individual capacity.

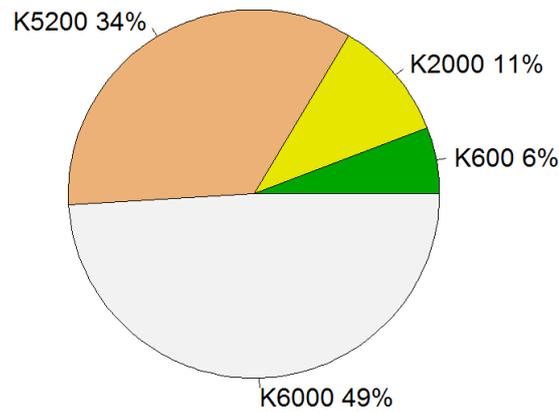

Fig. 7. Relative computing capacity.

For the test number #4, Figure 8 shows all the execution times of GPUs and relative performances in comparison to the entry-level K600 model.

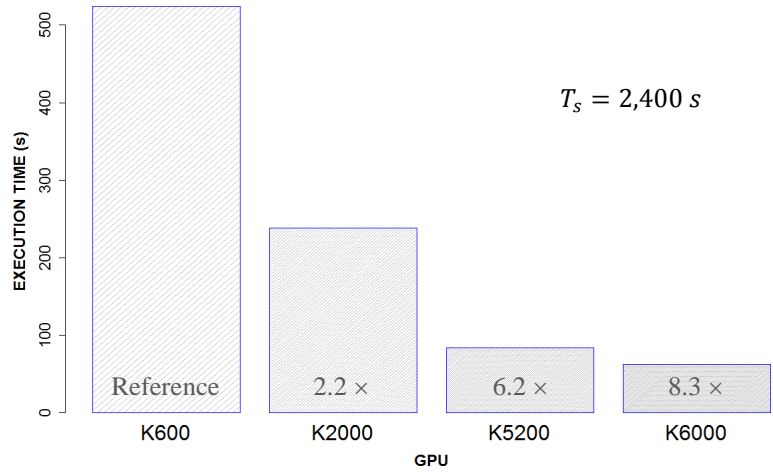

Fig. 8. GPU Execution times for test number #4.

K5200 and K6000 are the only ones GPUs able to load the problem test #5 in memory as it requires more than 2 GB of RAM. The performance of these two graphics cards are important for large grids as shown in Figure 9.

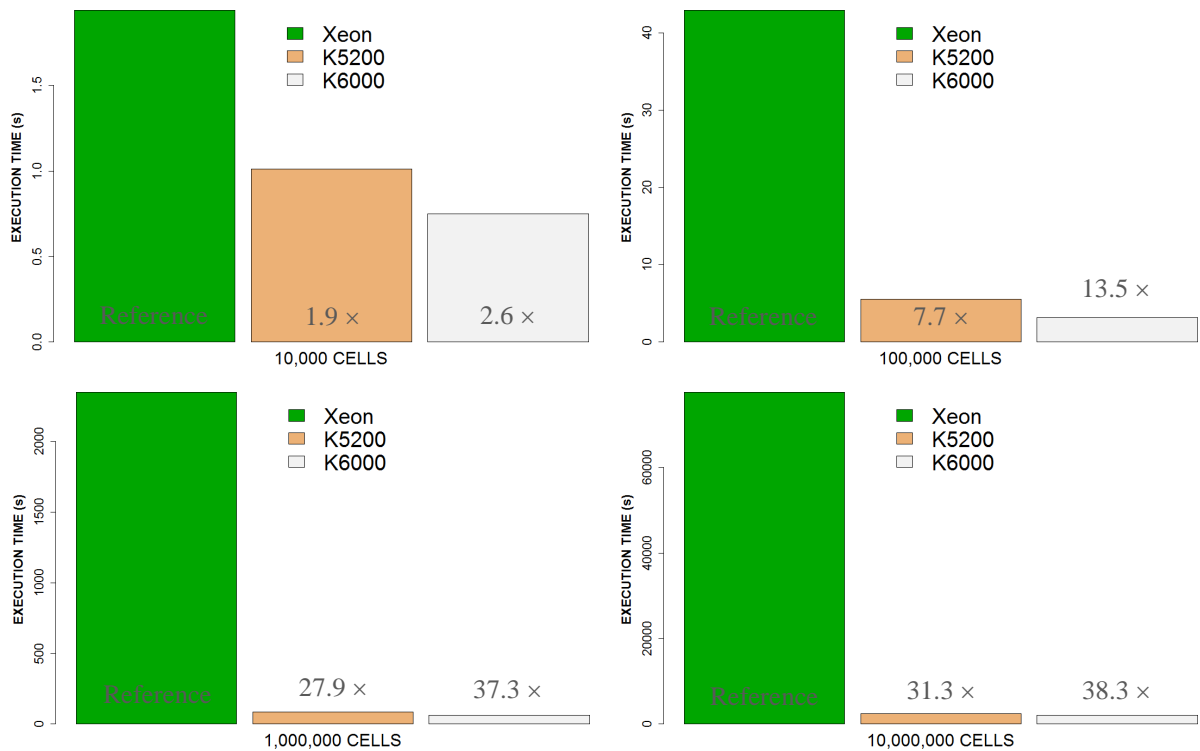

Fig. 9. Execution times for test number #2, #3, #4 and #5.

The speed-up can also depends on the simulation time $T_s$. Figure 10 shows for the model K6000, how the speed-up varies with the simulation time.

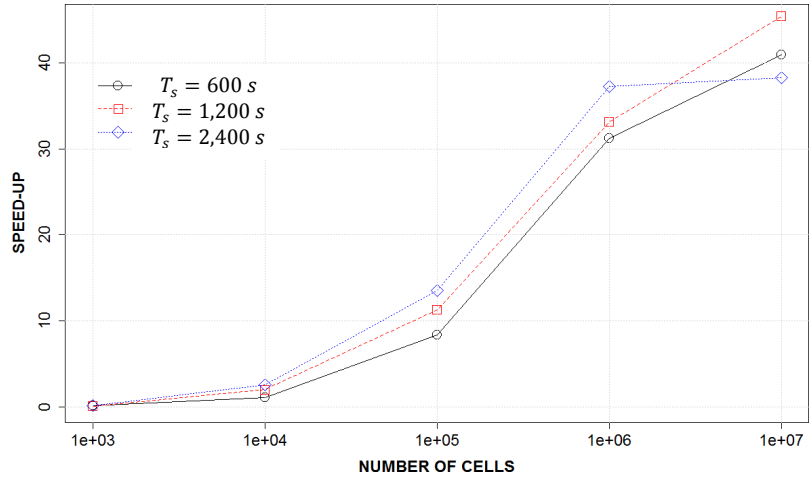

Fig. 10. Speed-up as a function of the simulation time.

## 4. Discussion

The exploitation of the parallel processing power of GPU is interesting in comparison to a single-core GPU if the problem size of the finite volume solver for shallow water systems is significant. This minimum size of interest lies in the range starting at 10,000 cells for high performance graphics card like the Quadro K5200/K6000 to 100,000 cells for the entry-level K600.

The maximum speed-up of the execution time is also bounded by the model of GPU. Its value is approximately equal to ×3 for the Quadro K600 to more than ×40 for K6000. This confirms that the performance highly depends on the model of the graphics card. Nevertheless, for all the simulations times and grids, the speed-up factor remains more dispersed for high-end GPUs if one takes a look at the statistical variability as shown in Figure 11.

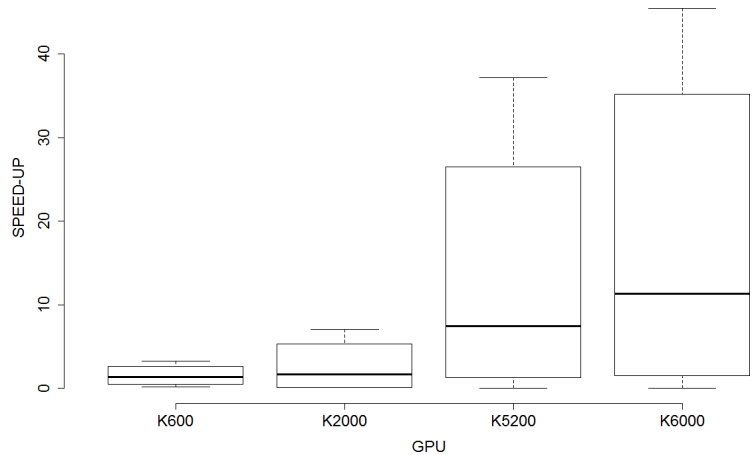

Fig. 11.Dispersion of results on the speed-up factor.

All the studied GPU in this paper come from the NVIDIA Quadro line and mainly differ by the following technical specifications: processing cores, RAM and memory bandwidth. Whereas the amount of RAM is important for dealing with large size problems, the number of processing cores and the rate at which data can be transferred to or from memory are important for the overall performance. The joint analysis of Table 1 and Figure 7 exhibits the direct relationship between performance and technical specifications, as shown in Figure 12.

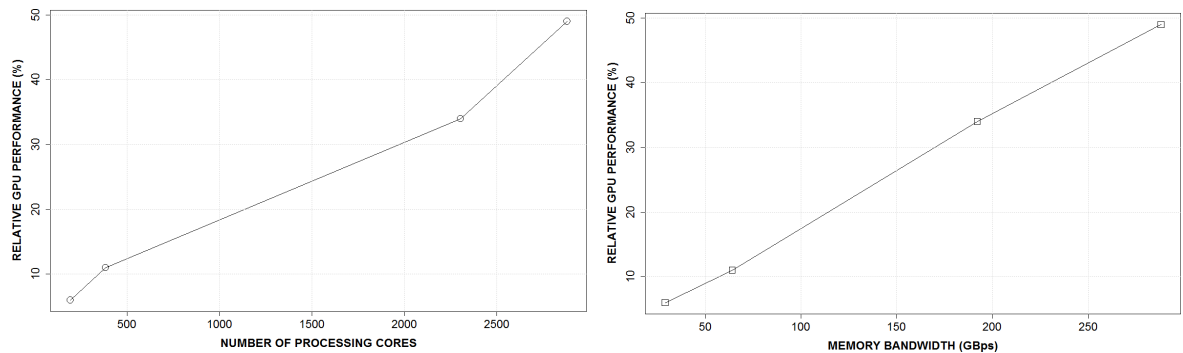
Fig. 12. GPU performance as a function of the technical specifications.

In Figure 10, the speed-up depends on the value of the simulation time with largest values for longest times except for one point. This behavior still needs to be investigated on a larger variety of problems in order to conclude.

## 5. Conclusions

This paper presents a parallelization of SWE for GPU under the NVIDIA CUDA framework. GPU computations are compared with a sequential approach on CPU. Several comparison tests are done involving different grid sizes, simulation times and GPU models. GPU can significantly speed-up the simulations according to the choice of computational parameters.

More tests still have to be done to investigate the influence of the simulation time on the speed-up. And finally, a logical continuation of this work would also compare the multi-GPU approach vs multi-CPU.